\pdfoutput=1
\documentclass[11pt]{article}

\usepackage[utf8]{inputenc}
\usepackage[T1]{fontenc}
\usepackage{lmodern}
\usepackage{ragged2e}
\usepackage{amsmath, amssymb, amsfonts}
\usepackage{graphicx}
\usepackage{float}
\usepackage{changepage}
\usepackage{hyperref}
\usepackage{geometry}
\title{Introducing the Correlation Concentration Ratio (CCR): 
Quantitative Framework for Comparing Quantum Cluster States}

\author{
  Amin Ahadi, Saman Sarshar \\
  ICQTs \\
  \texttt{saman.sarshar@shahroodut.ac.ir}
}

\date{}

\begin{document}

\maketitle

\begin{abstract}
In this paper, numerical simulations of four-mode continuous-variable cluster states with different topologies in the framework of measurement-based quantum computation are presented. By utilizing the symplectic representation and covariance matrix, the process of generating cluster states with linear, square, and T-shaped topologies has been systematically modeled. The simulation results show that the cluster graph structure is directly reflected in the pattern of quadrature correlations; in other words, the theoretical nullifier relations of the cluster states are reproduced in the final covariance matrices. Increasing the squeezing parameter leads to the strengthening of the target correlations and the suppression of unwanted components arising from anti-squeezing; such that the off-diagonal elements of the covariance matrix in the linear and square topologies increase to significant values, and in the T-shaped topology a stronger central correlation (similar to GHZ-like behavior in the continuous-variable domain) is observed. In order to quantitatively analyze these structural differences, a metric titled CCR (Correlation Concentration Ratio) is introduced that quantifies the concentration of effective correlations on the graph edges relative to the total correlations of the system. This index enables direct comparison of different topologies from the perspective of structural entanglement distribution and provides a framework for evaluating the efficiency of cluster graphs in MBQC architectures. The results show that CCR can be used as a practical tool for designing and selecting optimal topologies in larger clusters and more complex structures.
\end{abstract}

\setcounter{section}{0}

\section{Introduction}

Measurement-based quantum computation as one of the fundamental paradigms of quantum computation provides a conceptual alternative to the quantum circuit model. In this approach, instead of applying a sequence of quantum gates on input states, a pre-prepared multipartite entangled state is used and quantum information processing is performed solely through appropriate measurements on its subsystems. Cluster states are recognized as the primary resources for this type of computation and play an essential role in its realization.

Raussendorf and Briegel (2001) introduced the one-way quantum computer\cite{Raussendorf2001}. Menicucci et al. (2006) demonstrated universal quantum computation with continuous-variable cluster states\cite{ Menicucci2006}.

In the domain of continuous variables (CV), Gaussian cluster states have attracted significant attention due to their experimental producibility in optical systems, compatibility with scalable architectures, and the possibility of implementation based on optical squeezing. Recent advances in the generation of large-scale clusters in the time and frequency domains have also enabled fault-tolerant architectures. Larsen et al. (2021) proposed a fault-tolerant continuous-variable measurement-based quantum computation architecture\cite{Larsen2021}.

In the graphical framework of these states, each vertex corresponds to an optical mode and each edge represents a controlled-phase interaction between adjacent modes. Therefore, the topology of the cluster graph directly determines the pattern of entanglement and the path of quantum information propagation in the MBQC process. Studying different topologies, including linear, square, and T-shaped structures, can provide a deeper understanding of the role of graph geometry in computational efficiency. For example, star or T-shaped structures can reproduce GHZ-like behaviors in the CV domain, while two-dimensional topologies with higher symmetry are more suitable for scalable architectures. Yukawa et al. (2008) experimentally generated four-mode continuous-variable cluster states\cite{Yukawa2008}.

Since CV cluster states are generally Gaussian, their complete description can be performed using the covariance matrix. This matrix contains information about the quadrature variances and inter-mode correlations and enables precise analysis of the quantum features of the system without requiring the full Wigner function in the infinite-dimensional phase space. Therefore, the covariance matrix is an efficient tool for numerical examination of the entanglement structure and the quality of cluster nullifiers\cite{ Hao2021,González2021}.

For modeling the evolution of Gaussian states and applying quantum gates in this framework, the symplectic representation is a suitable tool. Symplectic transformations, rooted in classical mechanics, provide a mathematical structure under which the commutation relations of quadratures are preserved. In quantum optics, these transformations allow a simple and coherent representation of linear operations such as controlled-phase gates and directly affect the covariance matrix. Thus, the symplectic representation is an appropriate tool for accurate modeling of the cluster state generation process and their numerical analysis\cite{Booth2023}. 

In this paper, using the symplectic representation and covariance matrix analysis, numerical simulations of four-mode cluster states with linear, square, and T-shaped topologies have been performed. In addition to reproducing the theoretical nullifier relations and examining the dependence of correlation quality on the squeezing parameter, To the best of our knowledge, we introduce a quantitative index titled CCR (Correlation Concentration Ratio) is introduced. This metric is directly extracted from the inter-mode quadrature correlations and quantifies the structural concentration of correlations on the graph edges relative to the overall correlation distribution in the system. Unlike standard entanglement measures that only assess the existence or overall amount of entanglement, CCR is designed to reveal topology-dependent anisotropy in correlation distribution.a feature directly related to the manner of information propagation and computational stability in MBQC.

The presented results can be a step toward clarifying the role of graph geometry in the performance of measurement-based quantum computation and providing a quantitative framework for comparing different topologies in continuous-variable clusters)\cite{Du2023,Larsen2021,Hao2021}.

\section{Methods}
In this section, the results of numerical simulations of four-mode cluster states in linear, square, and T-shaped topologies are presented. The simulations were performed in the symplectic representation framework, and the complete description of the quantum states was carried out solely based on the covariance matrix. All operations applied to the initial states were of Gaussian type; therefore, the final states are also Gaussian. Consequently, the covariance matrix contains all the information required for analyzing the variances and quadrature correlations. In the following, in addition to the standard covariance-based analysis, the CCR index is introduced for quantitative comparison of the topologies.

\subsection{Graph Topology and Correlation Structure}
In this paper, numerical simulations of four-mode cluster states have been performed in the framework of measurement-based quantum computation in the continuous-variable domain. Given that the initial states and applied operations are Gaussian, the complete description of the system can be achieved solely based on the symplectic representation and covariance matrix. This approach enables precise analysis of the quantum features of cluster states without requiring the Wigner function in the infinite-dimensional phase space.

In the first step, the initial state of the system is considered as four independent modes, each in a momentum-squeezed vacuum state. The initial covariance matrix is defined as a block-diagonal matrix whose diagonal elements specify the position and momentum quadrature variances of each mode. These variances are determined in terms of the squeezing parameter, and simulations for different values of this parameter were performed to investigate the effect of squeezing level on correlation quality.

After defining the initial state, the necessary interactions for creating the cluster state were implemented by applying controlled-phase gates between adjacent modes. In the symplectic representation, each controlled-phase gate acts as a linear transformation on the quadrature vector and is described by a corresponding symplectic matrix. In this transformation, the position quadrature remains unchanged, and the momentum quadrature of each mode is shifted proportionally to the position quadrature of the neighboring mode.

For each topology, the corresponding graph adjacency matrix was first defined. This matrix plays a key role in constructing the overall symplectic matrix of the system, such that the connection information between modes is directly incorporated into the structure of this transformation. Using the adjacency matrix, the overall symplectic matrix was constructed as a block matrix acting on the quadrature vector of the entire system.

After determining the symplectic matrix, the final covariance matrix of the cluster state was calculated from the standard covariance transformation relation under symplectic transformations.

\[S^\top \sigma_{\text{initial}}\, S = \sigma_{\text{final}}\tag{1}
\]

This relation enables direct calculation of variances and quadrature correlations and forms the basis of the numerical analysis of cluster states. In this way, the quantum features of the final state can be extracted analytically–numerically without the need for time-domain dynamics simulation or random sampling\cite{González2021,Booth2023,Solodovnikova2025}.

To examine the dependence of cluster state quality on physical parameters, simulations were performed for different values of the squeezing parameter. The obtained covariance matrices were displayed graphically so that the pattern of correlations between quadratures could be directly observed. In these plots, diagonal elements represent the noise level of each quadrature, and off-diagonal elements indicate the quantum correlations between different modes.

\begin{figure}[H]
\centering
\includegraphics[width=0.85\textwidth]{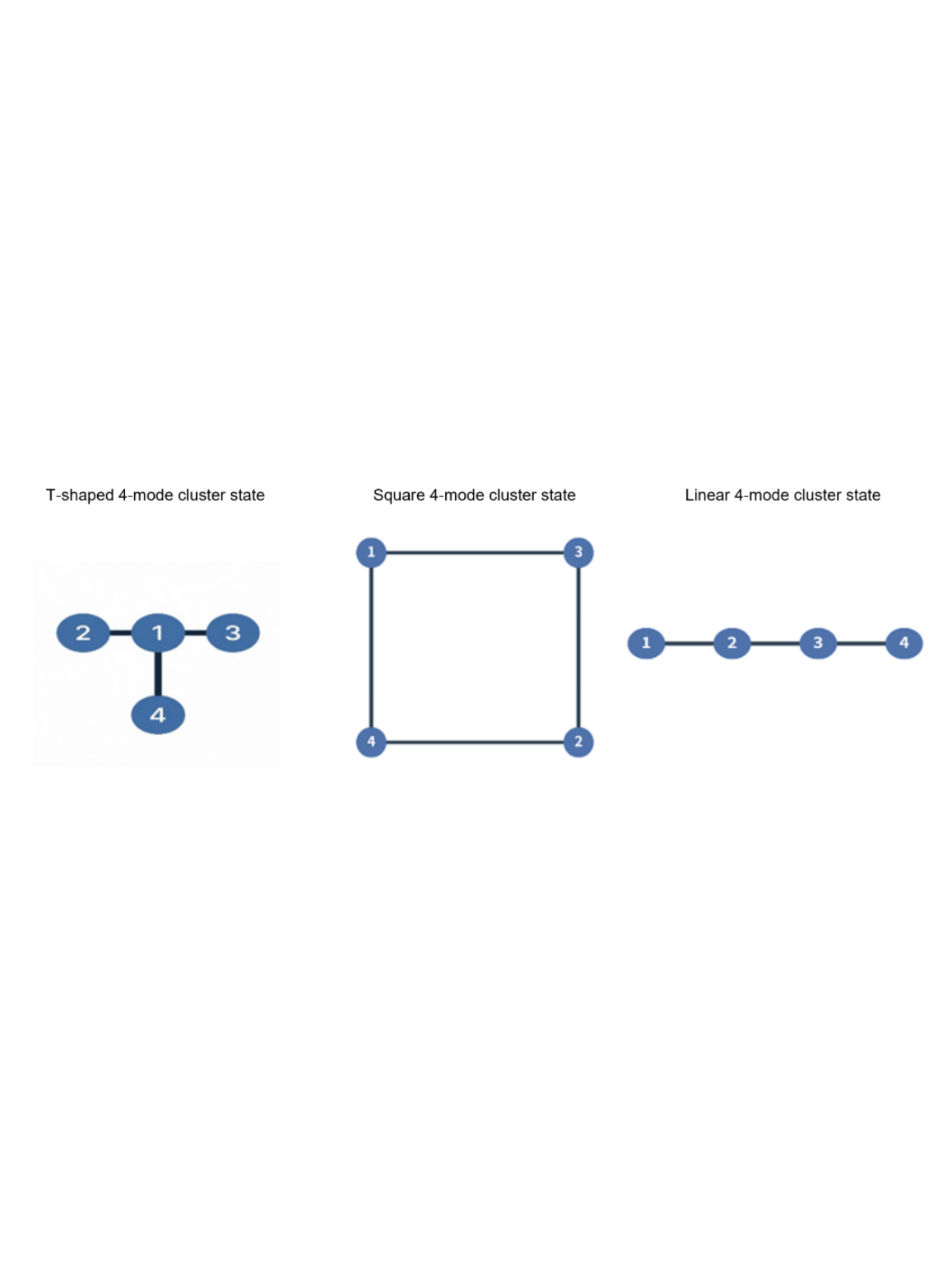}
\caption{schematic cluster for linear, square, and T-shaped cluster states }
\label{fig:2}
\end{figure}

This methodology was applied uniformly to linear, square, and T-shaped cluster states (whose schematics are shown in Figure \ref{fig:2}), and the differences between these states appear only in the structure of the adjacency matrix and, consequently, in the final symplectic matrix.

Thus, direct comparison of the effect of graph topology on the structure of quadrature correlations has been made possible, and the role of graph geometry in determining entanglement features has been examined numerically\cite{Eaton2022}.

\subsection{Cluster States}
One of the resources required for performing measurement-based quantum computation is source states. Graph states are used as source states in measurement-based quantum computation\cite{Raussendorf2001,Gu2009}.

A graph state is formed by arranging a set of vertices (nodes) and edges. To create a suitable source state for measurement-based quantum computation, a qumode must be assigned to each vertex and a controlled-phase operation to each edge. Each graph state can be represented by the following relation\cite{Menicucci2006}.

\[
G = (V, E)
\tag{2}
\]

In the above relation, the number of graph vertices or the number of qumodes used to construct the graph state is denoted by $V$, and the number of edges (or the number of controlled-phase operations) is also denoted by $E$. The degree of each node, which is the number of edges connected to it, can be represented using the following relation.

\[
O(v_j) = n \tag{3a}
\]
\[
v_j \in V \tag{3b}
\]
\[
n \in \mathbb{N} \tag{3c}
\]

In the discrete-variable representation for creating a graph state, the positive eigenstate of the first Pauli operator can be assigned to each vertex, and a controlled-phase operation is placed between every two neighboring vertices. With this idea, the overall representation of each graph state is expressed as follows.

\[
|G\rangle = \prod_{(a,b) \in E} \ Z^{(a,b)} |+\rangle^{\otimes V}
\tag{4}
\]

In the above relation, $V$ is the number of graph vertices and $E$ corresponds to the edges of graph $G$, written as a set of ordered pairs of edges between every two vertices of graph $G$. Since controlled-phase operators are diagonal operators, the order of applying these operators has no effect on state $G$.

A graph state is an entangled state and has the property that by performing a measurement on any of the qumodes, the entanglement between the other qumodes remains intact. The graph state is known as the primary resource for implementing measurement-based quantum computation\cite{Menicucci2006,Du2023}.

If the graph corresponding to a graph state is a regular lattice (grid) in one, two, or three dimensions, the created state is called a cluster state in the literature. To create cluster states in optical systems, each vertex of the graph is placed in a momentum-squeezed state, and a controlled-phase gate is applied between every two neighboring vertices\cite{Zhang2006,Booth2023}.

\subsection{Symplectic Representation}
Two paths can be used to study cluster states: the Wigner function and the symplectic representation. Since the aim of this article is to study  cluster states, there will be no non-Gaussian factor in the study and simulation. Moreover, studying cluster functions using the Wigner function in phase space requires integration over infinite intervals. In this research, instead of using the Wigner function representation, the symplectic representation has been used to study cluster states. The symplectic representation has the capability that, instead of integration in infinite space, matrices can be used to study these states (cluster states). Using the symplectic formulation, the evolutions related to states and quantum operators can be examined as follows\cite{Weedbrook2012,González2021}.

\[
\bar{r}' \to S \bar{r} + \vec{d}
\tag{5}
\]

Also, for the transformation of operators (and the covariance matrix) we will have:

\[ V' = S V S^T\tag{6}\]

The symplectic representation is borrowed from classical mechanics. Under these transformations, the Hamiltonian equations and the Poisson bracket remain invariant. In quantum optics, these transformations are also used to examine and study Gaussian states. Since in this research only a four-mode cluster state has been studied and simulated, we expect the resulting cluster state from the simulation to be a Gaussian state. Therefore, choosing symplectic transformations is an effective and appropriate choice for analyzing the simulation results\cite{Weedbrook2012,Menicucci2006,Solodovnikova2025}.

In studies related to quantum optics, for a matrix (element) belonging to the symplectic group, the following condition must hold:
\[ S^T \Omega S = \Omega \tag{7}\]

For an optical system with $N$ modes, in the above relation $\Omega$ will be displayed as follows:

\[
\Omega = \bigoplus_{k=1}^{N} \omega
\tag{8}
\]

\[
\omega = \begin{pmatrix} 0 & 1 \\ -1 & 0 \end{pmatrix}
\tag{9}
\]

In this research, work has been done on a cluster state with 4 modes; therefore, $\Omega$ for this work will be written as follows\cite{Adesso2004}:

\[
\Omega = \begin{pmatrix}
0 & 1 & 0 & 0 & 0 & 0 & 0 & 0 \\
-1 & 0 & 0 & 0 & 0 & 0 & 0 & 0 \\
0 & 0 & 0 & 1 & 0 & 0 & 0 & 0 \\
0 & 0 &  -1  & 0 & 0 & 0 & 0 \\
0 & 0 & 0 & 0 & 0 & 1 & 0 & 0 \\
0 & 0 & 0 & 0 & -1 & 0 & 0 & 0 \\
0 & 0 & 0 & 0 & 0 & 0 & 0 & 1 \\
0 & 0 & 0 & 0 & 0 & 0 & -1 & 0
\end{pmatrix}
\tag{10}
\]

\subsection{Covariance Matrix for Cluster States}
In continuous-variable measurement-based quantum computation, the covariance matrix is used to provide a complete Gaussian description of quantum states. This matrix is presented using position and momentum quadratures by the following relation:

\[
V_{ij} = \frac{1}{2} \langle\{\hat{\zeta}_i, \hat{\zeta}_j \} \rangle - \langle\hat{\zeta}_i \rangle\langle\hat{\zeta}_j \rangle
\tag{11}
\]

In the above relation, for each of the $zeta_{ij}$ we have the quadrature correlations.

\[
\hat{\zeta} = (\hat{x}_1, \hat{p}_1, \dots, \hat{x}_N, \hat{p}_N )^T
\tag{12}
\]

The diagonal elements of the covariance matrix determine the variance level of each quadrature; also, the off-diagonal elements of this matrix show the quadrature correlations (which can indicate the existence of entanglement) between the quadratures of different modes\cite{Simon2000,Weedbrook2012,González2021}.

In the process of preparing a four-mode cluster state, the initial covariance matrix can be written as follows. 

\[
V_0 = \begin{pmatrix}
\frac{1}{2} e^{+2r} I_4 & 0 \\
0 & \frac{1}{2} e^{-2r} I_4
\end{pmatrix}
\tag{13}
\]

Applying the controlled-phase gate on each connected edge between adjacent vertices affects as follows. 

\[
\vec{\hat{x}} \rightarrow \vec{\hat{x}}
\tag{14}
\]

\[
\vec{\hat{p}} \rightarrow \vec{\hat{p}} + A \vec{\hat{x}}
\tag{15}
\]

Also, the symplectic matrix related to this problem (four-mode cluster) can be displayed in block-diagonal form as follows\cite{Zhang2006,Booth2023,Menicucci2006}.

\[
S = \begin{pmatrix}
I_4 & 0 \\
A & I_4
\end{pmatrix}
\tag{16}
\]

Using this matrix, the covariance matrix after cluster creation can be obtained.

\[
V = S V_0 S^T
\tag{17}
\]

For this problem, according to the figure below, the adjacency matrix can be written as follows. 

\[
A = \begin{pmatrix}
0 & 1 & 0 & 0 \\
1 & 0 & 1 & 0 \\
0 & 1 & 0 & 1 \\
0 & 0 & 1 & 0
\end{pmatrix}
\tag{18}
\]

This matrix is used in creating the symplectic matrix\cite{Menicucci2006,Du2023,Solodovnikova2025,Zhang2006}.

\subsection{Numerical Simulation}
In this section, the figures related to the covariance matrices for each of the four-mode linear, square, and T-shaped clusters are presented in order. In these matrices (plots), diagonal elements represent quadrature variances and off-diagonal elements indicate quantum correlations between modes.
The simulation results for each of the linear, square, and T-shaped clusters are as follows.

The following results correspond to the linear cluster for Squeezing values ranging from 3 to 16 dB.
\begin{figure}[H]
\centering
\includegraphics[width=0.85\textwidth]{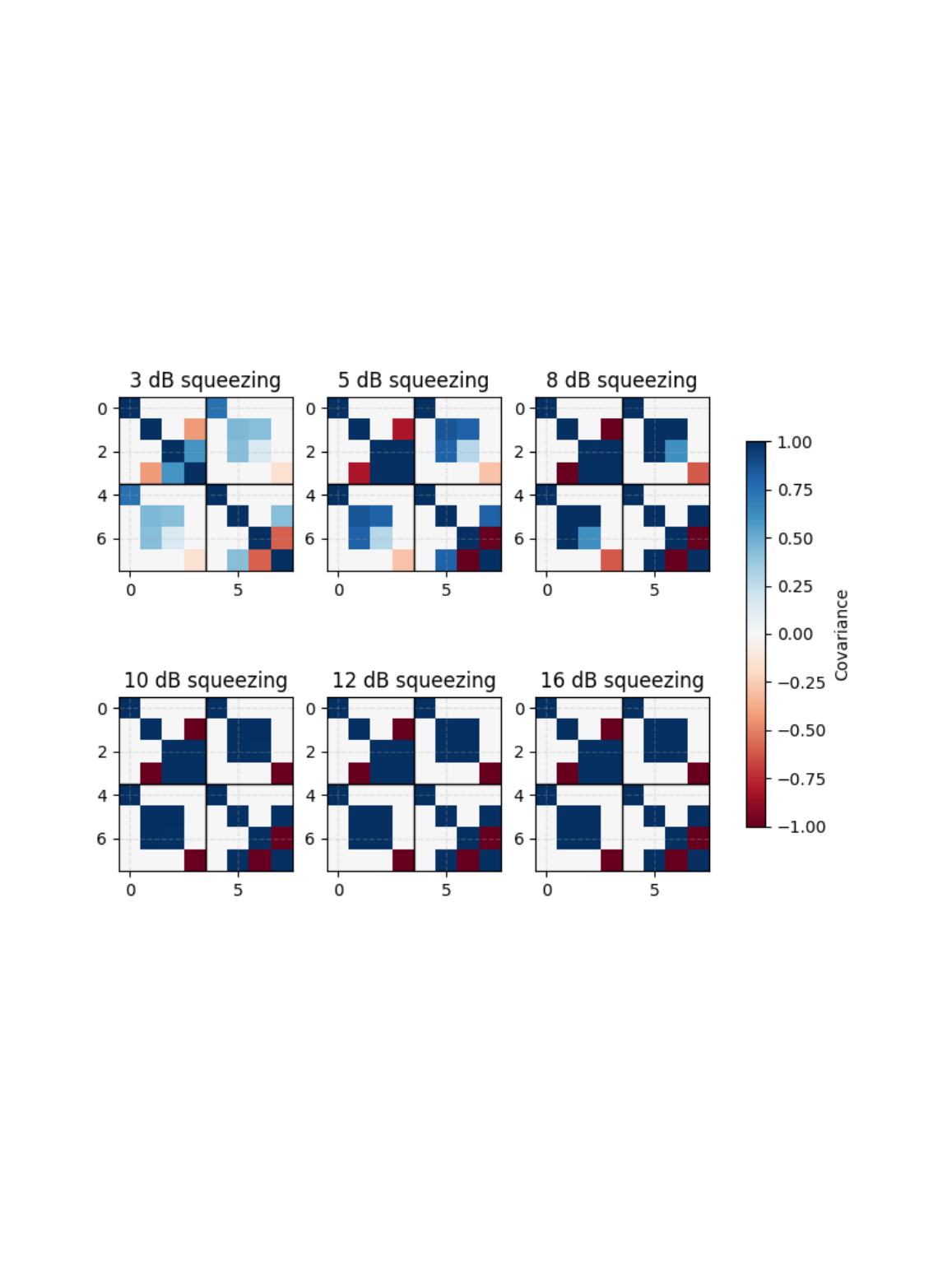}
\caption{Covariance matrix for the four-mode linear cluster with squeezing parameter 3-16dB}
\label{fig:1}
\end{figure}

The following results correspond to the square cluster for Squeezing values ranging from 3 to 16 dB.

\begin{figure}[H]
\centering
\includegraphics[width=0.85\textwidth]{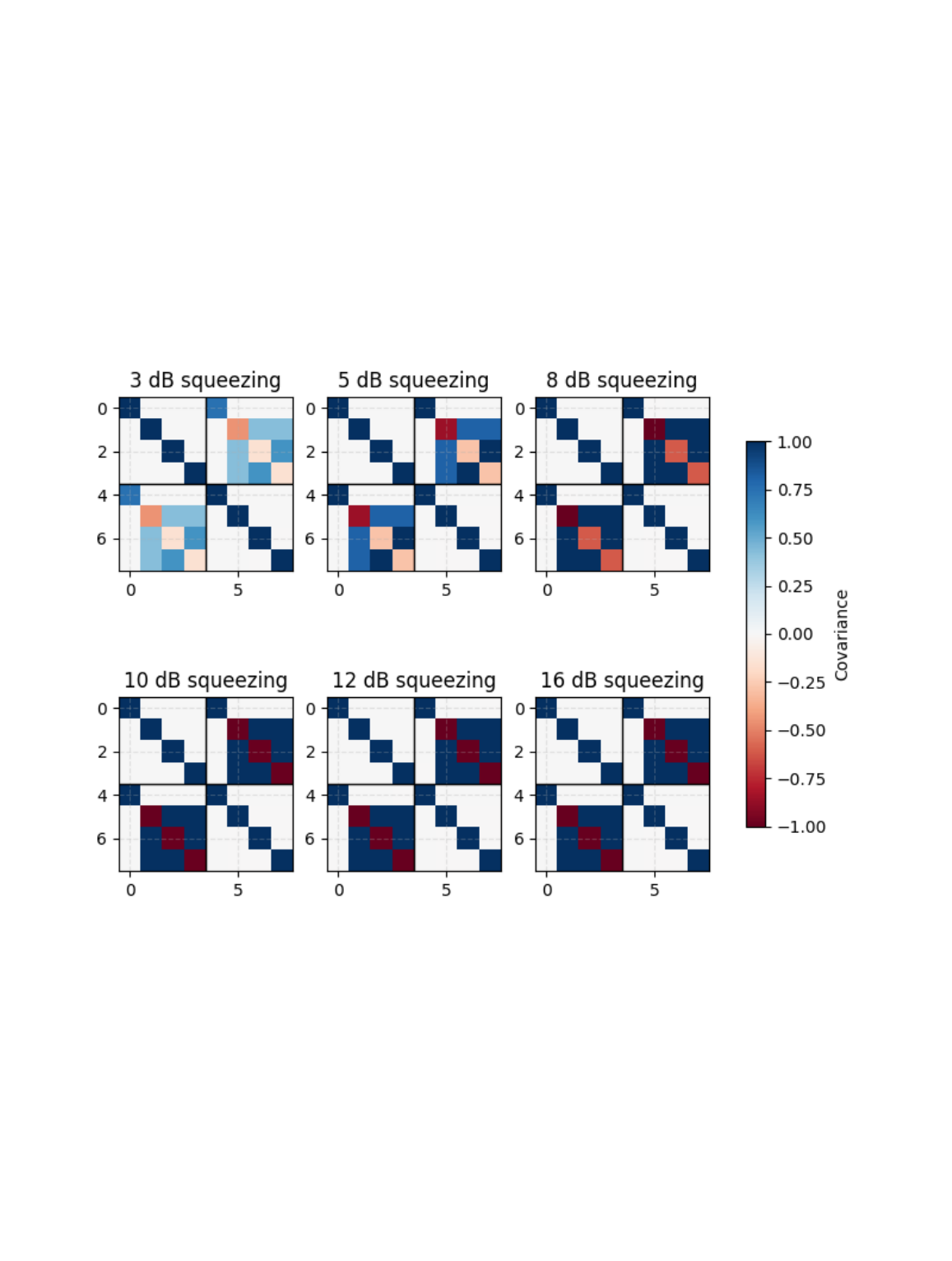}
\caption{Covariance matrix for the four-mode square cluster with squeezing parameter 3-16dB}
\label{fig:3}
\end{figure}

The simulation results for each of the linear, square, and T-shaped clusters are as follows.

The following results correspond to the T-shaped cluster for Squeezing values ranging from 3 to 16 dB.

\begin{figure}[H]
\centering
\includegraphics[width=0.85\textwidth]{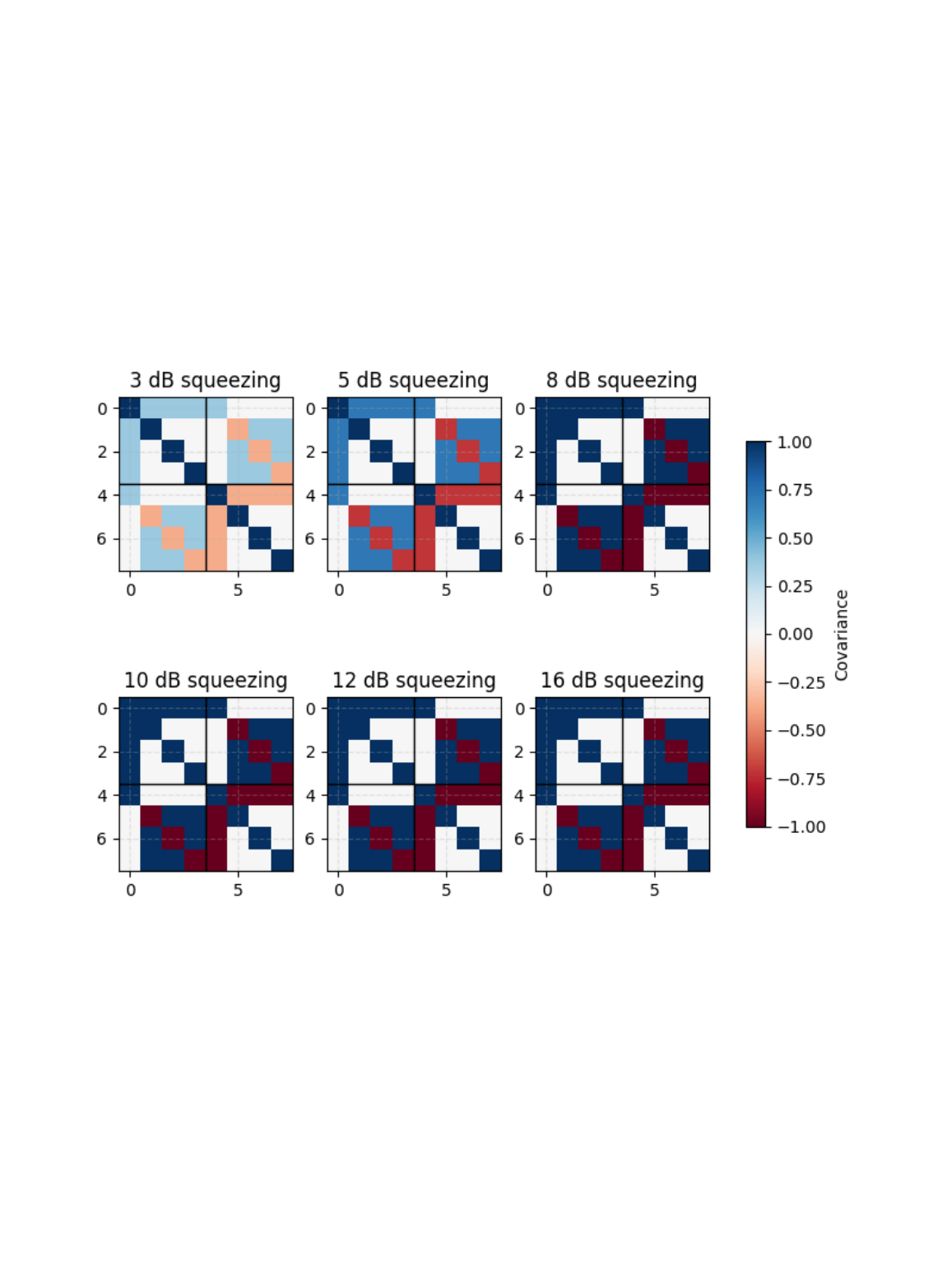}
\caption{Covariance matrix for the four-mode T-shaped cluster with parameter 3-16dB}
\label{fig:5}
\end{figure}

For the four-mode linear cluster state, the obtained covariance matrix indicates that off-diagonal elements corresponding to the correlations between the momentum quadrature of each mode and the position quadrature of neighboring modes exhibit non-zero and significant values. This correlation pattern is precisely consistent with the nullifier relations employed in the definition of ideal cluster states, where linear combinations of momentum and position quadratures, appearing in the theoretical definition of these states, possess reduced variances. Furthermore, components corresponding to undesired correlations, which might arise from anti-squeezing, have been effectively suppressed. This suggests that the symplectic transformation network has been designed correctly, and a high-quality linear cluster state has been produced.

As the squeezing parameter increases, the intensity of the quadrature correlations intensifies, consequently leading to larger values for the corresponding off-diagonal elements in the covariance matrix, while the overall structure of this matrix and the neighborly pattern among modes remain unchanged. This behavior indicates that the cluster graph topology is independent of the squeezing value, and only the quality of the entanglement generated depends on the amount of squeezing.

The results pertaining to the four-mode square cluster state show that its covariance matrix exhibits symmetric correlations between each mode and its two adjacent modes. These correlations are distinctly observed in the blocks corresponding to the position-momentum quadratures, and their pattern perfectly aligns with the structure of the square graph. Since the square cluster state is related to the linear cluster state through localized Fourier transforms, a considerable similarity is observed in the intensity of correlations between these two states, while the main difference lies in the distribution of these correlations among the modes. This outcome is fully consistent with theoretical predictions regarding the local equivalence of cluster states.

For the four-mode T-shaped cluster state, the covariance matrix obtained from the simulation reveals that the strongest correlations are established between the central mode and its three peripheral modes, while no significant direct correlation is observed between the peripheral modes themselves. This characteristic clearly reflects the T-shaped graph topology and conforms to the theoretical nullifier relations for this state. Such a correlation pattern confirms that the simulated state accurately reproduces the properties of a T-shaped cluster state. This structure is also consistent with interpreting this state as a GHZ-type state in the domain of continuous variables.

In summary, the analysis of the obtained covariance matrices demonstrates that the cluster graph topology is directly mirrored in the structure of the quadrature correlations, and the symplectic representation provides a suitable tool for studying these states. The simulation results are in complete agreement with the theoretical predictions for Gaussian cluster states, indicating that the employed method is capable of accurately generating and analyzing four-mode cluster states across various topologies.

\section{Correlation Concentration Ratio (CCR)}

In the context of continuous-variable Measurement-Based Quantum Computation (MBQC), the topology of the cluster graph dictates how quantum information propagates through the network. The quality of computation is not solely dependent on the degree of quantum correlations but also on their structural distribution among the modes. Inhomogeneities in the distribution of correlations can affect the pathways of information propagation and the manner in which errors spread within the system. Therefore, to quantify the topological structure of correlation distribution, we introduce an indicator known as the Correlation Concentration Ratio (CCR).

CCR is a numerical metric that describes the degree of concentration or distribution of inter-modal correlations relative to the cluster graph's structure. This indicator provides three main aspects:
1. Quantifying the uniformity or concentration of correlations throughout the cluster.
2. Offering a structural index for analyzing the directionality of information propagation in MBQC.
3. Providing a tool for assessing the sensitivity of the topology to perturbations or errors in specific modes.

Unlike qualitative approaches that rely solely on visual inspection of covariance matrices, CCR provides a numerical basis for comparing different topologies and selecting the appropriate structure for specific applications.

Generally, CCR is defined based on the off-diagonal elements of the final covariance matrix. If $V$ is the covariance matrix of the cluster state and $A$ is the graph's adjacency matrix, CCR can be defined as the ratio of the sum of the absolute values of correlations corresponding to the graph edges to the sum of all inter-modal correlations.

\[
\mathrm{CCR}
= \frac{
\displaystyle \sum_{i<j} A_{i,j}\left( |V_{x_i p_j}| + |V_{p_i x_j}| \right)
}{
\displaystyle \sum_{i<j} 
\left( |V_{x_i p_j}| + |V_{p_i x_j}| \right)
}
\tag{19}
\]

The Correlation Concentration Ratio (CCR) is defined in the canonical quadrature basis, which provides the natural representation for continuous-variable cluster states. Within this framework, the metric is restricted to cross-quadrature correlations, as these terms directly encode the entangling structure generated by controlled-phase interactions. To ensure that the measure captures the overall strength of correlations independently of their sign, absolute values are employed. This reflects the fact that CCR is intended to quantify correlation magnitudes rather than phase-dependent features. Furthermore, both $V_{x_i p_j}$ and $V_{p_i x_j}$ contributions are included in the definition in order to account for potential asymmetries that may arise in finite-squeezing regimes.

For different geometries, the numerator and denominator of this fraction can be calculated using their respective covariance matrices.
In this definition, $A_{ij}=1$ if an edge exists between modes $i$ and $j$, and $0$ otherwise. Therefore, CCR is a normalized quantity between zero and one, operating independently of the overall scale of correlations.

For the four-mode linear, square, and T-shaped topologies, CCR has been calculated using the specific structure of the adjacency matrix for each graph.Which is as follows for a linear topology as an example:

\[ 
\text{CCR}_{\text{Linear}} = 
\frac{
\displaystyle \sum_{(i,j) \in E_L} A_{i,j}\left( |V_{x_i p_j}| + |V_{p_i x_j}| \right)
}{
\displaystyle \sum_{i<j} 
\left( |V_{x_i p_j}| + |V_{p_i x_j}| \right)
}
\tag{20}
\]

 $E_L$, $E_T$, and $E_S$ denote the edge sets corresponding to the linear, T-shaped, and square cluster topologies, respectively.The plots illustrating the variations of the CCR within the squeezing range of 3 to 16 dB are presented in Fig.5.

\begin{figure}[H]
\centering
\includegraphics[width=\textwidth]{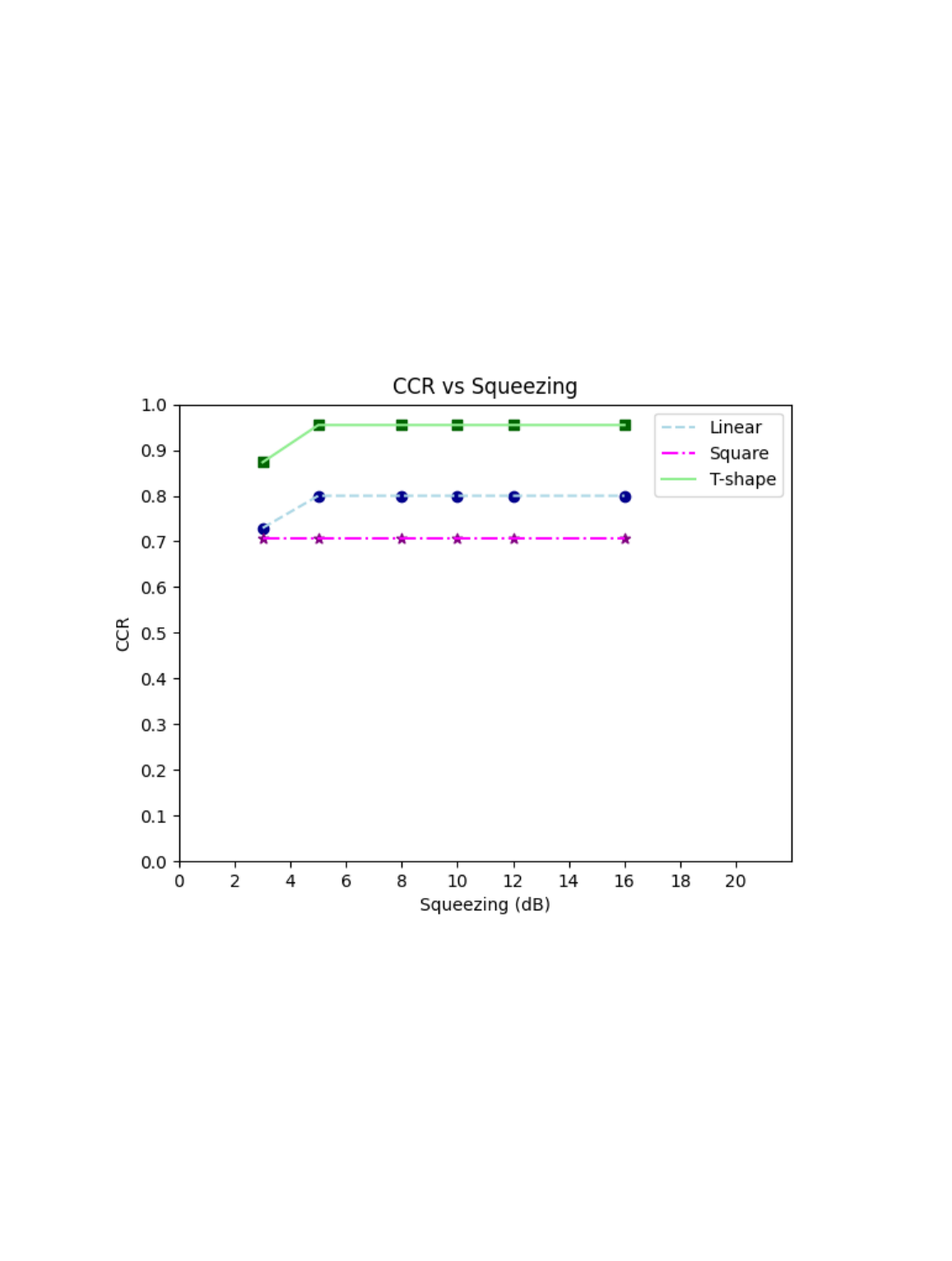}
\caption{CCR as a function of squeezing level (2–16 dB) for linear, square, and T-shaped cluster topologies.}
\label{fig:ccr}
\end{figure}

The CCR provides a quantitative measure describing how entanglement is distributed within graph-based structures.It is important to note that CCR does not quantify the total amount of entanglement, but rather its structural distribution relative to the underlying graph topology. The CCR values can be categorized into distinct regimes, each carrying a specific physical interpretation.

\textbf{Physical interpretation of CCR values}

\begin{itemize}

\item \textbf{smaller CCR values:}

smaller CCR values signifies a uniform distribution of correlations across the cluster. In a square topology (and generally topologies with distributed centrality), correlations are symmetrically spread among all modes. This feature implies the presence of redundant communication paths, which is essential for fault‑tolerant computation. In such structures, information can propagate through multiple distinct routes, and errors are symmetrically distributed. These values are commonly associated with geometries designed for universal and fault‑tolerant computation.

A smaller CCR value in this regime indicates that the centrality of the graph is distributed and that information travels through multiple paths, spreading symmetrically and omnidirectionally.

\item \textbf{Intermediate CCR values:}
Intermediate CCR values represent a semi‑centralized distribution of correlations, where certain modes play a dominant role in information transfer, leading to a more directional structure.

In a linear topology with intermediate CCR values, correlations are primarily concentrated along the chain, and the intermediate modes serve as information bottlenecks. This structure is suitable for sequential computation; however, any error occurring in the intermediate modes quickly propagates throughout the entire system.

In this regime, information spreads directionally and sequentially, making these structures more appropriate for computations with a prescribed ordered path.

\item \textbf{Larger CCR values:}

In this range, correlations become strongly concentrated on a single dominant mode. In a T‑shaped topology, the central mode acts as a quantum hub through which all communication between peripheral modes is routed. This structure provides complete isolation between peripheral modes but makes the central node a critical point of failure, increasing vulnerability to errors. This regime is mainly relevant in multi‑party communication settings where all information passes through the central mode.

Thus far, we have considered a 4-mode system. However, it is important to note that this formula remains applicable for a higher number of modes, allowing us to infer the suitability of various topologies for different applications based on their respective CCR values.

The CCR addresses \textit{where} and \textit{how} modes are distributed. What distinguishes CCR from other metrics is its focus on the topological structure of correlation distribution. This contrasts with metrics like QMI, which only measure \textit{how much} correlation exists. This distinction is critical for applications such as Measurement-Based Quantum Computation (MBQC), as the distribution of correlations directly impacts information propagation and error spreading. These factors, in turn, influence the choice of graph for a given problem.

The CCR is generalizable to a larger number of modes, with each topology exhibiting its own specific scaling behavior.

Topologies with a balanced, distributed centrality are the most suitable for scalability, as their CCR values change only slightly and remain small. These symmetric topologies inherently maintain a CCR small regardless of size, rendering them ideal for scalable architectures.

Topologies falling into the intermediate CCR category exhibit increasing inhomogeneity as the number of modes increases. This typically occurs in linear topologies, indicating the growing dominance of intermediate states with increasing chain length.

In the third category of topologies, which we now refer to as star-shaped (a generalized N-state extension of the T-shaped cluster) instead of T-shaped, the concentration intensifies with an increasing number of modes. Star-shaped topologies exhibit super-linear growth with N, as the central node accumulates correlations with an increasing number of peripheral states, while peripheral-to-peripheral correlations remain negligible.

This scaling analysis indicates that square topologies are the most scalable for universal MBQC, as their CCR remains balanced with increasing system size—a critical feature for fault-tolerant quantum computation\cite{Raussendorf2005}.

.

To investigate the dependence of CCR on the amount of squeezing, we consider a range from 3 to 16 dB. This range covers from the minimum experimentally observable values to those approaching practical fault-tolerance thresholds. The results indicate that although the intensity of correlations grows with increasing squeezing, the relative behavior of CCR remains predominantly a function of topology.

Although CCR is not a standard graph-theoretic quantity, its scaling behavior reflects the uniformity of the distribution of quadrature correlations, which is directly related to the homogeneity of resources required for fault-tolerant MBQC.

As observed in the figure above, topologies with CCR small, which includes the symmetric distribution, exhibit the least variation with increasing squeezing levels. This indicates that they are the optimal choice for fault tolerance.

\section*{Comparison of CCR with Other Measures of Quantum Correlations}

Several metrics exist for quantifying quantum correlations. Quantum Mutual Information (QMI) measures the total correlations of a system but lacks direct dependence on the topological structure of the graph. Negativity is a measure of bipartite entanglement, computable for Gaussian states, but it does not provide information about how correlations are structurally distributed. Quantum Discord also quantifies the deviation from classicality but depends on the choice of subsystems\cite{Adesso2005,Cerf1997,Ollivier2002,Henderson2001,Moraes2024,Laurell2024,Modi2012,Chen2024}.

In contrast, CCR is a structural-topological measure directly derived from the covariance matrix. Its focus lies on the distribution of correlations relative to the cluster graph. This indicator complements standard metrics rather than replacing them, offering distinct information related to graph geometry that is crucial for MBQC analysis.

\end{itemize}

\section{Conclusion}
In this work, a numerical simulation of four-mode cluster states within the Continuous-Variable Measurement-Based Quantum Computation (CV-MBQC) framework was presented. Utilizing the symplectic representation and covariance matrix-based analysis, the structure of quadratures’ correlations was investigated for linear, square, and T-shaped topologies. The results demonstrated that the theoretically nullifying relations are accurately reproduced within the covariance structure, and the dependency of correlation intensities on the squeezing parameter aligns with theoretical predictions.

The most significant achievement of this work is the introduction of the CCR index as a structural quantity for analyzing the topological distribution of correlations. This index enables a quantitative comparison of different topologies, revealing that the graph’s geometry plays a decisive role in how quantum correlations are concentrated and propagate. The presented framework is scalable to larger clusters and can serve as a foundation for future analyses concerning scalability and noise modeling in CV-MBQC architectures.



\begin{thebibliography}{99}

\bibitem{Raussendorf2001} Raussendorf\& Briegel, H. J. A One-Way Quantum Computer. \textit{Phys. Rev. Lett.} \textbf{86}, 5188--5191 (2001).
\bibitem{Menicucci2006} Menicucci\textit{et al.} Universal Quantum Computation with Continuous-Variable Cluster States. \textit{Phys. Rev. Lett.} \textbf{97}, 110501 (2006).
\bibitem{Larsen2021} Larsen, M. V. \textit{et al.} Fault-Tolerant Continuous-Variable Measurement-Based Quantum Computation Architecture. \textit{PRX Quantum} \textbf{2}, 030325 (2021).
\bibitem{Yukawa2008} Yukawa, M. \textit{et al.} Experimental generation of four-mode continuous-variable cluster states. \textit{Phys. Rev. A} \textbf{78}, 012301 (2008).
\bibitem{Hao2021} Hao, S. \textit{et al.} Quantum computation and error correction based on continuous variable cluster states. \textit{Chin. Phys. B} \textbf{30}, 060312 (2021).
\bibitem{González2021} González, C. \textit{et al.} Cluster States from Gaussian States: Essential Diagnostic Tools for Continuous-Variable One-Way Quantum Computing. \textit{PRX Quantum} \textbf{2}, 030343 (2021).
\bibitem{Booth2023} Booth\& Markham, D. Flow conditions for continuous variable measurement-based quantum computing. \textit{Quantum} \textbf{7}, 1146 (2023).
\bibitem{Du2023} Du\textit{et al.} Generation of large-scale continuous-variable cluster states multiplexed both in time and frequency domains. \textit{Opt. Express} \textbf{31}, 7535--7544 (2023).


\bibitem{Eaton2022} Eaton, M. \textit{et al.} Measurement-based generation and preservation of cat and grid states within a continuous-variable cluster state. \textit{Quantum} \textbf{6}, 769 (2022).
\bibitem{Gu2009} Gu, M. \textit{et al.} Quantum computing with continuous-variable clusters. \textit{Phys. Rev. A} \textbf{79}, 062318 (2009).
\bibitem{Solodovnikova2025} Solodovnikova \& Neergaard-Nielsen, J. S. Fast simulations of continuous-variable circuits using the coherent state decomposition. Preprint at https://arxiv.org/abs/2508.06175 (2025).

\bibitem{Zhang2006} Zhang\& Braunstein, S. L. Continuous-variable Gaussian analog of cluster states. \textit{Phys. Rev. A} \textbf{73}, 032318 (2006).
\bibitem{Weedbrook2012} Weedbrook, C. \textit{et al.} Gaussian quantum information. \textit{Rev. Mod. Phys.} \textbf{84}, 621--669 (2012).
\bibitem{Adesso2004} Adesso, G., Serafini, A. \& Illuminati, F. Extremal entanglement and mixedness in continuous variable systems. \textit{Phys. Rev. A} \textbf{70}, 022318 (2004).
\bibitem{Simon2000} Simon, R. Peres-Horodecki separability criterion for continuous variable systems. \textit{Phys. Rev. Lett.} \textbf{84}, 2726 (2000).
\bibitem{Raussendorf2005} Raussendorf, R., Harrington, J. \& Goyal, K. A fault-tolerant one-way quantum computer. \textit{Ann. Phys.} \textbf{321}, 2242--2270 (2006).


\bibitem{Cerf1997} Cerf, N. J. \& Adami, C. Quantum mutual information. \textit{Phys. Rev. Lett.} \textbf{79} (1997).
\bibitem{Ollivier2002} Ollivier, H. \& Zurek, W. H. Quantum discord. \textit{Phys. Rev. Lett.} \textbf{88}, 017901 (2002).
\bibitem{Henderson2001} Henderson, L. \& Vedral, V. Classical and quantum correlations. \textit{J. Phys. A} \textbf{34}, 6899--6905 (2001).
\bibitem{Moraes2024} Moraes, G. L., Angelo, R. M. \& Costa, A. C. S. Axiomatic approach to measures of total correlations. \textit{Entropy} \textbf{26}, 238 (2024).
\bibitem{Laurell2024} Laurell, P. \textit{et al.} Witnessing entanglement and quantum correlations in condensed matter. \textit{Adv. Quantum Technol.} \textbf{7} (2024).
\bibitem{Modi2012} Modi, K. \textit{et al.} The classical-quantum boundary for correlations. \textit{Rev. Mod. Phys.} \textbf{84}, 1655--1707 (2012).

\bibitem{Adesso2005} Adesso, G. \& Illuminati, F. Gaussian measures of entanglement versus negativities: the ordering of two-mode Gaussian states. \textit{Phys. Rev. A} \textbf{72}, 032334 (2005).
\bibitem{Chen2024} Chen, X. Scalable multipartite entanglement criteria for continuous variables. Preprint at https://arxiv.org/abs/2411.03083 (2024).

\end{thebibliography}
\end{document}